\documentclass[reqno]{article}
\usepackage{amssymb,amsmath,epsfig,cite}

\begin{document}
\begin{center}{\large \bf On spinless null propagation in five dimensional space-times with approximate space-like Killing symmetry}
\vskip.25in
Romulus Breban\footnote{E-mail address: breban@gmail.com}

{\it Institut Pasteur, 75724 Paris Cedex 15, France} %
\end{center}
\marginparwidth 0pt \oddsidemargin 0pt \evensidemargin 0pt

\begin{abstract} 
Five-dimensional (5D) space-time symmetry greatly facilitates how a 4D observer perceives the propagation of a single spinless particle in a 5D space-time. In particular, if the 5D geometry is independent of the fifth coordinate then the 5D physics may be interpreted as 4D quantum mechanics. In this work we address the case where symmetry is approximate, focusing on the case where the 5D geometry depends weakly on the fifth coordinate. We show that concepts developed for the case of exact symmetry approximately hold when other concepts such as decaying quantum states, resonant quantum scattering and Stokes drag are adopted, as well. We briefly comment on the optical model of the nuclear interactions and Millikan's oil drop experiment. 

\end{abstract}


\section{Introduction}
\label{sec:intro}

Geodesic propagation in five dimensional (5D) space-times remains, lately, a topic of great interest \cite{Diemer:2014fw,Chandler:2015ip,Grunau:2013bg,Grunau:2013fw,Guha:2012kw,Leon,Seahra:2001vi,Wesson:2007ta}.  Particular attention has be devoted to 5D space-times, with a space-like fifth dimension which is neither compact nor Planckian, where the metric is independent of the fifth dimension \cite{Kleidis:1999nr,PhysRevD.38.1741,Seahra:2001vi,Wesson:2007ta}. It has been shown that, in this case, 5D null-path propagation may be interpreted as {\it anomaly-free} 4D propagation both in the classical and quantum regimes \cite{Seahra:2001vi,Wesson:2007ta,Breban:2005a,Breban:2015a}. 
  
Why do we need thus a 5D formulation of physics? The approach to answer this question has been three fold. First, studies of field equations (e.g., see \cite{Wesson:2007ta,Kleidis:1999nr,PhysRevD.38.1741}) and geodesic motion \cite{Wesson:2007ta,Leon} claimed that the existence of extra dimensions manifests as new physics in ordinary 4D space-time. The relationship between geodesic motion and the classical tests of general relativity adapted for 5D space-times may be found in Ref.~\cite{Wesson:2007ta}.  Furthermore, it has been postulated that phase experiments may carry signature of the 5D nature of space-time \cite{Wesson:2015jz}.  

Second, it has been proposed that 5D space-time offers a unifying anthropic principle, since, inherently, we are 4D perceivers of Nature. In particular, it has been postulated that 5D symmetric space-times provide physical pictures depending on the symmetry \cite{Breban:2005a,Breban:2015a}. Hence, if the 5D geometry is independent of the fifth coordinate then the 5D physics can be perceived as 4D quantum mechanics, while if the 5D geometry is independent of time then the 5D physics can be perceived as 4D statistical mechanics.

Third, studies began to address general 5D space-time geometries.  A discussion of geodesic motion in non-symmetric 5D space-times is found in Ref.~\cite{Leon}.  In this work we follow the same line of thought and address the case where symmetry is approximate. We postulate how concepts developed for the case where the symmetry is exact are used for the case where the symmetry is approximate. Of course, these concepts will work for a small region where the 5D space-time is appears symmetric. However, we show that they can be extended over larger regions if additional concepts such as {\it decaying quantum states}, {\it resonant quantum scattering} and {\it Stokes drag} are adopted, as well. The price to pay for all these concepts working together is general covariance which no longer holds.  We focus on the case where the 5D geometry depends {\it weakly} on the fifth coordinate. The case where the 5D geometry depends weakly on time leads to the statistical physics of time-dependent systems, near thermodynamical equilibrium. We defer this to further work. 

The structure of the paper is as follows. In section \ref{sec:spacetime} we introduce our space-time geometry.  In section \ref{sec:quantum} we discuss quantum propagation using 5D null-path integrals.  In section \ref{sec:classic} we discuss geodesic motion and then we conclude our work.

\section{Five-dimensional space-time}
\label{sec:spacetime}

We consider a 5D space-time with the metric $h_{AB}$ ($A,B, ... =0,1,2,3,5$) containing one particle, and postulate that all quantum propagation of the particle takes place on 5D null paths \cite{Breban:2005a}.  A complete description of the propagation in the 5D space-time requires measurements of additional entities than 4D space-time events.  

We discuss a special class of 5D space-times, foliated into conformally-flat 4D space-times.  The conformal factor is chosen to be the inverse square lapse of the foliation. The shift of the foliation, $N_\mu\equiv qA_\mu/c^2$ ($\mu, \nu, ...=0,1,2,3$), remnant of the 5D gravitational field, is interpreted as an {\it external} electromagnetic field by the 4D observer. The symbol $q$ stands for the specific charge of the particle---we use cgs units throughout.  In this context, $q/c^2$ represents a conversion constant between the gravitational and the electromagnetic field. The 5D field equations imply that $A_\mu$ (i.e., $N_\mu$) satisfy Maxwell-type equations \cite{Wesson:2007ta,Kleidis:1999nr,PhysRevD.38.1741}. 

We thus obtain a flat 4D space-time with superimposed electromagnetic field, fit for describing many experimental setups. We think of these space-times as local approximations of more realistic geometric constructs including non-trivial gravitational fields, whose metrics satisfy suitable field equations \cite{Wesson:2007ta,Breban:2005a}.  Removing the conformal factor from the metric since it is irrelevant for null-path counting, we have
\begin{eqnarray}
\label{eq:fol1}
\tilde h_{AB}=\left(\begin{array}{cc}
\eta_{\mu \nu}+\frac{q^2}{c^4} A_\mu A_\nu & \frac{q}{c^2} A_\mu \\
              \frac{q}{c^2} A_\nu   & 1
\end{array}\right),\quad \tilde h^{AB}=\left(\begin{array}{cc}
\eta^{\mu \nu}& -\frac{q}{c^2} A^\mu \\
              -\frac{q}{c^2} A^\nu   & 1+\frac{q^2}{c^4} A^\rho A_\rho 
\end{array}\right),
\end{eqnarray}
where the indices $\mu, \nu, ...$ are raised with $\eta^{\mu\nu}$.  

\section{Path integrals}
\label{sec:quantum}
Consider any two causally ordered events 1 and 2, with 1 in the past of 2, which we write as $1\prec 2$. Denote the coordinates of 1 and 2 by $x^A_{(1)}$ and $x^A_{(2)}$, respectively. Then, the sum over 5D null paths between 1 and 2, denoted here by ${\mathcal R}(x^A_{(1)}, x^A_{(2)})$, is positively defined, conformally invariant, and has the status of a microcanonical sum, determining the particle propagation between 1 and 2 \cite{Breban:2005a}. The resulting 4D space-time is non-Lorentzian. The 5D infinitesimal null path element $ds_5\equiv\sqrt{\tilde h_{AB}dx^Adx^B}=0$ can be rewritten as 
\begin{eqnarray}
ds_{4\;\pm}\equiv dx^5=\pm\sqrt{-\eta_{\mu \nu}dx^\mu dx^\nu}-\frac{q}{c^2}A_\rho dx^\rho.
\end{eqnarray} 
If $A_\rho$ is independent of $x^5$, then we obtain a 4D quantum mechanical picture \cite{Breban:2005a}. $ds_{4\;\pm}$ can be regarded as a 4D metric in a 4D non-Lorentzian curved manifold, where the distance is $ds_{4\;\pm}$ if $1\prec 2$, and $-ds_{4\;\pm}$ if $2\prec 1$. Computing the path integral ${\mathcal R}(x^A_{(1)}, x^A_{(2)})$ in the 5D Lorentzian manifold with the metric $\tilde h_{AB}$ is equivalent to computing the following Feynman path integral \cite{Breban:2005a} 
\begin{eqnarray}
\Psi_{\pm}(\lambda^{-1};x^\mu_{(1)}, x^\mu_{(2)})=\int\limits[d^4x]e^{i\lambda^{-1}s_{4\;\pm}(1\prec 2)}
\end{eqnarray}
of paths between $x^\mu_{(1)}$ and $x^\mu_{(2)}$, the 4D projections of $x^A_{(1)}$ and $x^A_{(2)}$, respectively, in a non-Lorentzian manifold. The symbol $\lambda$ represents the Compton wavelength of the propagating particle; i.e., $\lambda\equiv\hbar/(mc)$. It follows that computing the path integral ${\mathcal R}(x^A_{(2)}, x^A_{(1)})$ is equivalent to computing $\Psi^*_{\pm}(\lambda^{-1};x^\mu_{(1)}, x^\mu_{(2)})$. $\Psi_{\pm}$ does not result as a scalar field on the 4D manifold, since a transformation of coordinates that reverses causality implies a complex conjugation of $\Psi_{\pm}$. 

If $A_{\rho}$ is not independent of $x^5$, then the traditional quantum mechanics picture no longer holds.  The mass of the particle is no longer a constant of motion. We discuss a possible interpretation of the case where we have weak electromagnetic fields of the form $A_\rho+x^5A'_\rho$ with $\partial_5A_\rho=0$ and $\partial_5 A'_\rho=0$. We obtain 
\begin{eqnarray}
\label{eq:x'^5}
dx^5&=&\pm\sqrt{-\eta_{\mu \nu}dx^\mu dx^\nu}-\frac{q}{c^2}A_\rho dx^\rho-\frac{q}{c^2}x^5A'_\rho dx^\rho.
\end{eqnarray}
Hence, it is not possible to naturally define a 4D path element. In fact, by separating the fifth coordinate and integrating, we obtain a non-local action 
\begin{eqnarray}
\label{eq:s'^5}
s'_{4\;\pm}(1\prec 2)\equiv x^5_{(2)}-x^5_{(1)}&=&\mkern-14mu\;\int\limits_{1\prec 3 \prec 2}\mkern-14mu\;\left( \pm\sqrt{-\eta_{\mu \nu}dx^\mu dx^\nu}-\frac{q}{c^2}A_\rho dx^\rho\right)\exp\left(\;\int\limits_{3\prec 2}dx^\rho\frac{q}{c^2}A'_\rho\right),
\end{eqnarray}
whose path integral
\begin{eqnarray}
\Psi'_{\pm}(\lambda^{-1};x^\mu_{(1)}, x^\mu_{(2)})=\int\limits_{1\prec 2}[d^4x]e^{i\lambda^{-1}s'_{4\;\pm}(1\prec 2)}
\end{eqnarray}
satisfies the following differential equation obtained according to Feynman's procedure \cite{Feynman:1948a} 
\begin{eqnarray}
\pm\frac{\hbar}{i}\frac{\partial \Psi'_\pm}{\partial t}&=&\frac{1}{2m}
\left[\frac{\hbar}{i}\nabla-\frac{mq}{c}\overrightarrow{A}\right]
^2\Psi'_\pm\mp mqA_0\Psi'_\pm+ mc^2\Psi'_\pm\nonumber\\ &~&+\frac{\hbar}{i}\left(-\frac{c}{2}\frac{q}{c^2}A'_0\pm\frac{q}{2mc^2}\overrightarrow{A'}\frac{\hbar}{i}\nabla\pm\frac{c}{2}\frac{q}{c^2}\overrightarrow{A}\frac{q}{c^2}\overrightarrow{A'}\right)\Psi'_\pm.
\end{eqnarray}
$\Psi'_-$ describes the forward propagation of a particle of mass $m$, while $\Psi'_+$ describes the forward propagation of an antiparticle of mass $m$. Since the electromagnetic field depends on $x^5$, mass is, in fact, no longer a constant of motion. However, the propagation is considered for a 4D particle of mass $m$, where terms in $A'_\rho$ account for the change in the dynamics due to the change in mass. 

The term in round brackets is non-hermitian and problematic for an interpretation of quantum mechanics where probability is conserved. Still, non-hermitian hamiltonians are used to describe key processes such as decaying quantum states \cite{1971PhRvA...4.1782M} and resonant quantum scattering \cite{1963PhRv..130..712B}. In particular, we emphasize the success of the phenomenological model of nuclear interactions in the range of 10-100 MeV known as the {\it optical model} \cite{Feshbach:1958a}. Summation over paths in space-times with complex coordinates, involving actions with complex potential, proved successful in further developing some of these theories \cite{Koeling:1975ub}.  Later, it was shown that paths in manifolds with real coordinates suffice\cite{1984PhLA..104..119W}. Our approach uses both real interaction potentials and paths in manifolds with real coordinates.

\section{Geodesic propagation}
\label{sec:classic}
We discuss the classical limit where $\lambda\rightarrow 0$ and the propagation takes place on a single path where the action achieves local extremum and which contributes most to the path integral. We denote by $\tau$ the parameter of a path and, using Eq.~\eqref{eq:s'^5}, write the lagrangian of the non-local action $s'_{4\;\pm}(1\prec 2)$, with $1\prec 3\prec 2$, as follows
\begin{eqnarray}
{\mathcal L}\equiv L+L' \mkern-14mu\;\int\limits_{1\prec 4\prec 3}\mkern-14mu\;d\tau L\exp\left(\;\int\limits_{4\prec 3}L'd\sigma\right),
\label{eq:LLL}
\end{eqnarray}
where
\begin{eqnarray}
L=\pm\sqrt{-\eta_{\mu \nu}\dot x^\mu \dot x^\nu}-\frac{q}{c^2}A_\rho \dot x^\rho,\quad L'=-\frac{q}{c^2}A'_\rho \dot x^\rho,
\end{eqnarray}
and $\dot x^\mu\equiv dx^\mu\slash d\tau$. Assuming that on the classical path 
\begin{eqnarray}
\label{eq:smallL'}
\left|\;\int\limits_{1\prec 2}L'd\sigma\right|\ll1,
\end{eqnarray}
we have\footnote{An alternate way to arrive at Eq.~\eqref{eq:LLL'} is as follows. Equation \eqref{eq:x'^5}, providing the infinitesimal action $dx^5$ as a function of $x^5$, may be read as an equation for an attracting fixed point at $x^5$. Hence, for the first order in $L'$, we perform one iteration toward the fixed point. That is, we choose $x^5_{(1)}=0$ and use $ds_{4\;\pm}(1\prec 3)$ for a nested approximation of $ds'_{4\;\pm}(1\prec 2)$.}
\begin{eqnarray}
\label{eq:LLL'}
{\mathcal L}\approx L+L' \int\limits_{1\prec 3}d\tau L,
\end{eqnarray}
and the equations Euler-Lagrange are
\begin{eqnarray}
\label{eq:EL}
\left[\frac{d}{d\tau}\left(\frac{\partial L}{\partial\dot x^\mu}\right)-\frac{\partial L}{\partial x^\mu}\right]+\nonumber\\
\left[\frac{d}{d\tau}\left(\frac{\partial L'}{\partial \dot x^\mu}\right)\int_{\tau_1}^\tau Ld\sigma+\frac{dL'}{d\tau}\int_{\tau_1}^\tau\frac{\partial L}{\partial x^\mu}d\sigma-\frac{\partial L'}{\partial x^\mu}\int_{\tau_1}^\tau Ld\sigma-L'\int_{\tau_1}^\tau\frac{\partial L}{\partial x^\mu}d\sigma+\frac{\partial L'}{\partial \dot x^\mu}L+L'\frac{\partial L}{\partial \dot x^\mu}\right]=0.
\end{eqnarray}

\subsection{A non-relativistic application}
To illustrate the 4D interpretation of 5D geodesic motion, we consider Eq.~\eqref{eq:EL} under the following conditions. First, we assume a simple form of the electromagnetic potential 
\begin{eqnarray}
\label{eq:EM}
A_\rho+x^5A'_\rho=(-zE,0,0,0)+x^5(A'_0,0,0,0), 
\end{eqnarray}
where $E$ and $A'_0$ are constants. Second, the path parameter $\tau$ represents the proper time of the time-like geodesic in the 4D manifold with the metric $\eta_{\mu\nu}$; i.e., $d\tau^2=-\eta_{\mu \nu}dx^\mu dx^\nu$. Third, we take the non-relativistic limit where $\tau\rightarrow ct$ and $|d\dot x^j d\dot x_j/c^2|\ll 1$. Fourth, we multiply Eq.~\eqref{eq:EL} by $mc$.\footnote{Because the translational symmetry along $x^5$ holds only approximately, $\lambda$ changes slowly with the proper time. We consider $\lambda$ as an {\it approximate} constant of motion. As a precaution for the physical interpretation, we use $\lambda^{-1}s'_{4\;\pm}$ rather than $s'_{4\;\pm}$ as physical action.}
Hence, we obtain 
\begin{eqnarray}
\label{eq:equmotion}
m\ddot z-\frac{mq}{c}A'_0\dot z\pm(mq)E\left[1-\frac{q}{c}A_0'(t_2-t_1)\right]=0,
\end{eqnarray}
where we further neglect the term $|(q/c)A_0'(t_2-t_1)|=|\int_{t_1}^{t_2}L'dt|\ll1$ [c.f.,~Eq.~\eqref{eq:smallL'}], meaning that the equation of motion holds for short time intervals $(t_2-t_1)\ll |c/(qA'_0)|$. Hence, we arrive at 
\begin{eqnarray}
\label{eq:equmotion2}
m\ddot z-\frac{mq}{c}A'_0\dot z\pm(mq)E=0.
\end{eqnarray}
This corresponds to the 5D picture whereby a particle with momentum $p^A=(E/c,\overrightarrow{p},mc)$ propagates with increasing mass, until eventually achieving the momentum $\hat p^A=(E/c,0,\hat mc)$, at rest. Equation \eqref{eq:equmotion2} stands for the motion of a 4D particle in constant electric field, subject to drag of type Stokes. In the case where $(mq)A_0'\slash c<0$, the 4D particle with speed $\dot z$ brakes because of interaction with a still enviroment. The case where $(mq)A_0'\slash c>0$ is reduced to the previous case by a time reversal transformation. However, an intrepretation by which the 4D particle is entrained by a moving environment is also posible.

The 5D setup with the electromagnetic potential given by Eq.~\eqref{eq:EM} has yet another interpretation \cite{Breban:2005a}. Since $A_\rho+x^5A'_\rho$ is time independent and motion is non-relativistic, the particle propagation may be described, as leading order, by the Langevin equation \cite{Breban:2005a}
\begin{eqnarray}
\label{eq:Lange}
m\ddot z+\zeta\dot z\pm(mq)E+\eta(t)=0,
\end{eqnarray}
where $\zeta$ is the drag coefficient and $\eta(t)$ is a stochastic force with $\langle\eta(t)\eta(t')\rangle=2\zeta k_BT\delta(t-t')$. In the limit where thermal fluctuations go to zero (i.e., $T\rightarrow 0$), the stochastic force vanishes. Hence, c.f., Eqs.~\eqref{eq:equmotion2} and \eqref{eq:Lange}, we identify $-(mq)A'_0\slash c$ with $\zeta$. Furthermore, we rewrite the condition $|(q/c)A_0'(t_2-t_1)|=|\int_{t_1}^{t_2}L'dt|\ll1$ as $2(t_2-t_1)\slash u\ll 1$, where $u\equiv 2m/\zeta$ is a {\it quantum of physical time} \cite{Breban:2005a}, defined for the case where $E=0$. 

Particle motion described by equations \eqref{eq:equmotion2} and \eqref{eq:Lange} was studied experimentally by Brown \cite{Brown:1866a} and Millikan \cite{Millikan:1911p1030,Millikan:1913va}. In particular, Millikan studied the motion on charged oil drops in constant electric field. By tuning $E$, while keeping everything else unchanged, one could in principle measure $mq$. However, to estimate $mq$, Millikan observed oil drops whose charge changed suddenly during the experiment, due to ionization. Hence, according to our theory, this corresponds to several {\it single particle} propagations. Explaining the experimental results obtained by Millikan requires additional physical principles than those described here.

\section{Discussion and conclusion}
Kaluza's ansatz to set up $\partial_5$ to zero leads to a unified formulation of field equation for the gravitational and electromagnetic fields \cite{Kaluza:1921}. However, even in the case where the metric is $x^5$-dependent, one may absorb the terms which contain $\partial_5$ into sources of the electromagnetic field. Hence, formally, a split of the 5D field equations between equations for the gravitational and the electromagnetic fields is possible even in the absence of the space-like symmetry.  Therefore, we conclude that Maxwell-type equations, in agreement to experimental observations of electric fields, may be written even though the 5D space-time is not symmetric.

In previous work, we discussed an interpretation of a symmetric 5D space-time and introduced physical concepts such as mass and temperature to 4D geometry. In the case where the 5D space-time is approximatively symmetric, these concepts hold approximatively, in a small region of space-time. The description of the 5D geometry may be improved by using additional concepts such as decaying states, resonant quantum scattering and Stokes drag. These concepts correct the previously proposed 4D picture \cite{Breban:2005a} and extend the description of particle propagation for longer durations of proper time. Furthermore, we link newly obtained equations of motion to previously studied phenomenology. Hence, building on previous work, more and more 4D physics may be regarded from the perspective of a non-compact 5D space-time.


\begin{thebibliography}{10}
\ifx \showCODEN  \undefined \def \showCODEN #1{CODEN #1}  \fi
\ifx \showISBN   \undefined \def \showISBN  #1{ISBN #1}   \fi
\ifx \showISSN   \undefined \def \showISSN  #1{ISSN #1}   \fi
\ifx \showLCCN   \undefined \def \showLCCN  #1{LCCN #1}   \fi
\ifx \showPRICE  \undefined \def \showPRICE #1{#1}        \fi
\ifx \showURL    \undefined \def \showURL {URL }          \fi
\ifx \path       \undefined \input path.sty               \fi
\ifx \ifshowURL \undefined
     \newif \ifshowURL
     \showURLtrue
\fi

\bibitem{Diemer:2014fw}
Valeria Diemer, Jutta Kunz, Claus L{\"a}mmerzahl, and Stephan Reimers.
\newblock {Dynamics of test particles in the general five-dimensional
  Myers-Perry spacetime}.
\newblock {\em Physical Review D}, 89\penalty0 (12):\penalty0 124026, June
  2014.

\bibitem{Chandler:2015ip}
Jesse Chandler and Moataz~H Emam.
\newblock {Geodesic structure of five-dimensional nonasymptotically flat
  2-branes}.
\newblock {\em Physical Review D}, 91\penalty0 (12):\penalty0 125024, June
  2015.

\bibitem{Grunau:2013bg}
Saskia Grunau and Bhavesh Khamesra.
\newblock {Geodesic motion in the (rotating) black string spacetime}.
\newblock {\em Physical Review D}, 87\penalty0 (12):\penalty0 124019, June
  2013.

\bibitem{Grunau:2013fw}
Saskia Grunau, Valeria Kagramanova, and Jutta Kunz.
\newblock {Geodesic motion in the (charged) doubly spinning black ring
  spacetime}.
\newblock {\em Physical Review D}, 87\penalty0 (4):\penalty0 044054, February
  2013.

\bibitem{Guha:2012kw}
Sarbari Guha, Pinaki Bhattacharya, and Subenoy Chakraborty.
\newblock {Particle motion in the field of a five-dimensional charged black
  hole}.
\newblock {\em Astrophysics and Space Science}, 341\penalty0 (2):\penalty0
  445--455, 2012.

\bibitem{Leon}
J~Ponce de~Leon.
\newblock Equations of motion in kaluza-klein gravity reexamined.
\newblock {\em Gravitation and Cosmology}, 8:\penalty0 272--284, 2002.

\bibitem{Seahra:2001vi}
S~S Seahra and P~S Wesson.
\newblock {Null geodesics in five-dimensional manifolds}.
\newblock {\em General Relativity and Gravitation}, 33\penalty0 (10):\penalty0
  1731--1752, 2001.

\bibitem{Wesson:2007ta}
P~S Wesson.
\newblock {\em {Space-time-matter}}.
\newblock Modern Higher-dimensional Cosmology. World Scientific, 2007.

\bibitem{Kleidis:1999nr}
K~Kleidis and D~Papadopoulos.
\newblock {On the adiabatic expansion of the visible space in a
  higher-dimensional cosmology}.
\newblock {\em General Relativity and Gravitation}, 29\penalty0 (3):\penalty0
  275--290, March 1997.

\bibitem{PhysRevD.38.1741}
Luis Faria-Busto.
\newblock Some new cosmological results of quadratic lagrangians.
\newblock {\em Phys. Rev. D}, 38:\penalty0 1741--1753, Sep 1988.
\newblock \ifshowURL {\showURL
  \path|http://link.aps.org/doi/10.1103/PhysRevD.38.1741|}\fi.

\bibitem{Breban:2005a}
R~Breban.
\newblock Interpretation of the five dimensional quantum propagation of a
  spinless massless particle.
\newblock {\em Progress of Theoretical Physics}, 114\penalty0 (3):\penalty0
  643--668, 2005.

\bibitem{Breban:2015a}
R~Breban.
\newblock A five-dimensional perspective on the klein--gordon equation.
\newblock {\em Ann Phys}, 356:\penalty0 158--170, 2015.

\bibitem{Wesson:2015jz}
Paul~S Wesson.
\newblock {Extra dimensions and phase experiments}.
\newblock {\em Annalen der Physik}, 528\penalty0 (3-4):\penalty0 307--312,
  November 2015.

\bibitem{Feynman:1948a}
Richard~P. Feynman.
\newblock Space-time approach to non-relativistic quantum mechanics.
\newblock {\em Reviews of Modern Physics}, 20:\penalty0 367--87, 1948.

\bibitem{1971PhRvA...4.1782M}
R~M More.
\newblock {Theory of Decaying States}.
\newblock {\em Physical Review A}, 4\penalty0 (5):\penalty0 1782--1790,
  November 1971.

\bibitem{1963PhRv..130..712B}
B~Buck.
\newblock {Calculation of Elastic and Inelastic Proton Scattering with a
  Generalized Optical Model}.
\newblock {\em Physical Review}, 130\penalty0 (2):\penalty0 712--726, April
  1963.

\bibitem{Feshbach:1958a}
H~Feshbach.
\newblock The optical model and its justification.
\newblock {\em Annual Reviews of Nuclear Science}, \penalty0 (49-104), 1958.

\bibitem{Koeling:1975ub}
Thijs Koeling and R~A Malfliet.
\newblock {Semi-classical approximations to heavy ion scattering based on the
  Feynman path-integral method}.
\newblock {\em Physics Reports}, 22\penalty0 (4):\penalty0 181--213, 1975.

\bibitem{1984PhLA..104..119W}
E~M Wright.
\newblock {Path integral approach to the Schr{\"o}dinger equation with a
  complex potential}.
\newblock {\em Physics Letters A}, 104\penalty0 (3):\penalty0 119--122,
  December 1983.

\bibitem{Brown:1866a}
Robert Brown.
\newblock {\em The miscellaneous botanical works of Robert Brown}, volume~1.
\newblock R. Hardwicke, 1866.

\bibitem{Millikan:1911p1030}
RA~Millikan.
\newblock {The isolation of an ion, a precision measurement of its charge, and
  the correction of Stokes's law.}
\newblock {\em Phys. Rev.}, 32\penalty0 (4):\penalty0 349, 1911.

\bibitem{Millikan:1913va}
RA~Millikan.
\newblock {Brownian Movements in Gases at Low Pressures}.
\newblock {\em Phys. Rev Phys Rev}, 1\penalty0 (3):\penalty0 218--221, 1913.

\bibitem{Kaluza:1921}
T~Kaluza.
\newblock {Zum Unit{\"a}tsproblem der Physik}.
\newblock {\em Sit. Preuss. Akad. Wiss. Phys. Mat.}, k1:\penalty0 966, 1921.

\end{thebibliography}
\end{document}